\documentclass[conference]{IEEEtran}

\usepackage[utf8]{inputenc}
\usepackage{amsmath}
\usepackage{graphicx}
\usepackage{xcolor}
\usepackage{tikz}
\usepackage{pgfplots}



\makeatother

\usepackage{stfloats}

\begin{document}
%
\title{Robust Homomorphic Video Hashing}
\author{\IEEEauthorblockN{Priyanka Singh}
\IEEEauthorblockA{Dhirubhai Ambani Institute of Information and Communication Technology\\
Gandhinagar, Gujarat, India\\
Email: Priyanka\_Singh@daiict.ac.in}}

\maketitle

\begin{abstract}
The Internet has been weaponized to carry out cybercriminal activities at an unprecedented pace. The rising concerns for preserving the privacy of personal data while availing modern tools and technologies is alarming. End-to-end encrypted solutions are in demand for almost all commercial platforms. On one side, it seems imperative to provide such solutions and give people trust to reliably use these platforms. On the other side, this creates a huge opportunity to carry out unchecked cybercrimes.  
 This paper proposes a robust video hashing technique, scalable and efficient in chalking out matches from an enormous bulk of videos floating on these commercial platforms. The video hash is validated to be robust to common manipulations like scaling, corruptions by noise, compression, and contrast changes that are most probable to happen during transmission. It can also be transformed into the encrypted domain and work on top of encrypted videos without deciphering. Thus, it can serve as a potential forensic tool that can trace the illegal sharing of videos without knowing the underlying content. Hence, it can help preserve privacy and combat cybercrimes such as revenge porn, hateful content, child abuse, or illegal material propagated in a video. 
\end{abstract}

\IEEEpeerreviewmaketitle

\section{Introduction}

The unprecedented growth in the rate of multimedia capture and transmission technologies is witnessing a massive explosion of data worldwide. According to recent statistics in 2017, every minute of YouTube upload accounts to nearly 300 hours of video upload. The average number of visitors browsing YouTube everyday accounts to nearly 30 million. Facebook, the biggest social media site, has a daily average of 8 billion video views from around 500 million users. 

Given the massive growth of online videos, retrieval of efficient videos is a challenging task. The design of a fast and efficient content-based search for videos is very challenging compared to images. A video has a complex, diverse pattern of high-level information across the frames like action recognition, event detection, etc. apart from the low-level information contained within the frames \cite{Chou2015PatternBasedNV, Hoi2008AMA,hu2011survey,kofler2014intent,yu2016scalable}. Even a small snippet of a video needs massive computational resources to extract a signature that can be used for mapping across the vast repository of the available videos. The development of an efficient framework that can facilitate fast retrieval of relevant videos for a given query video is a challenging research problem.

Videos are more susceptible to security and privacy breaches as they carry both absolute and contextual information. They may lead to serious consequences if the video carries any confidential information. Upcoming news about privacy losses has already alarmed society. Hence, end-to-end encryption based solutions are in high demand. Maintaining the utility of the solutions and addressing security and privacy concerns is a challenging research area.

Homomorphic encryption is a class of encryption schemes that allows operations on the encrypted content but is computationally quite expensive. Catering the problem of the videos with this homomorphic encryption tool becomes even more challenging. With so much bulk content in a video, the hash extraction must be designed very strategically to be feasibly used in real-time scenarios. 


In this paper, we describe a secure video hashing algorithm that extracts a unique signature from the encrypted videos exploiting the homomorphic properties of the Pallier cryptosystem. The signature is robust against common manipulations like compression quality, additive noise, re-scaling, and non-linear intensity adjustments (gamma correction). The algorithm is designed based on the restricted homomorphic operations, i.e., addition and scalar multiplication, so that the same signature can be extracted from its corresponding unencrypted (plaintext) version. Large-scale experiments are performed to validate the efficacy of the algorithm.
  

The rest of the paper is organized as follows: the related research on video hashing is summarized in Section \ref{sec:related}. Section \ref{sec:paillier} discusses in brief the homomorphic properties of Paillier cryptosystem, Section \ref{sec:videohash} discusses the proposed video hashing and then its transformation in the encrypted domain is discussed in Section \ref{sec:encrypted}. The validation of the proposed algorithm is briefed in Section \ref{sec:exp}, followed by discussion in Section \ref{sec:discussion}.

\section{Related Work}
\label{sec:related}


Videos contribute a significant proportion to the deluge of data shared every day on social media sites. Though substantial contribution has been made to content-based image retrieval methods in the literature, video hashing is still very limited. This is largely contributed to the complex nature of the video contents. The broad spectrum of information ranges from low-level features contained in the individual frames to high-level semantic-based information. There can be diversified dynamic backgrounds and complex cues of people's activities. It becomes incredibly challenging to perform a similarity-based search given the enormous video corpus on the Web.

Traditional video hashing schemes mainly aimed to address two major categories of problems: near-duplicate retrieval and content-based video retrieval (CBVR).  For the former class, conventional hashing techniques were employed for efficient identification of duplicate videos \cite{coskun2006spatio, douze2010image, song2013effective}. For the second class, most semantically similar videos were searched for a given query video \cite{cao2012submodular, li2015hierarchical, ye2013large, yu2016scalable}. 

CBVR schemes can be further divided into three sub-categories: Schemes that extract the video information in the form of a representative feature vector and then design efficient hashing functions to work on top of it. The second class extends the image hashing algorithm to videos by treating each frame of video as an image and computes a similarity score by averaging the individual frame-to-frame distance metrics. However, this straightforward application is quite computationally expensive, as even a small snippet of a video has more than fifty frames considering an average frame rate of 20 fps. The third category of schemes first selects some frames as representative of video information and then applies the image hashing algorithm to these frames to extract a hash value. This is quite challenging as a trade-off needs to be maintained between retaining the necessary informing and throwing away any repeated information to keep up the time efficiency of the hash. Even these schemes are unable to exploit the temporal information of the videos.

Song et al. \cite{song2013effective} employed multiple feature extraction methods that exploited the local structures of the frames to obtain the binary codes. Douze et al. \cite{ douze2010image} used the traditional hashing algorithms to extract the binary codes.  Cao et al. \cite{cao2012submodular} extracted a few representative frames and used a sub-modular hashing framework to derive a hash value that was used for video search. However, no statistical heuristics were used in this scheme to arrive at the video hash. A linear hash function which exploited the temporal information of the videos via employing the supervised structural learning was proposed by Ye et al. \cite{ye2013large}.  A  video retrieval scheme exploiting the kernel max-margin spanning through the Euclidean space and the Riemannian manifold was proposed by Li et al. \cite{li2014compact, li2015hierarchical}. Though they represented features through a covariance matrix, the Spatio-temporal information remains to be exploited.

Hence, it is desirable to present a video hashing framework that learns strong features by utilizing the temporal information, and in turn, also minimizes the hamming distance computation. The design of an efficient distance metric to retrieve the most relevant videos based on a query video is a very challenging problem. Unlike searches based on textual titles that are easier to implement employing an inverted index structure, proposing a similarity metric for videos is challenged by the difficulty of modeling the video semantics.
      
      Here, we present a robust video hashing algorithm that is computationally very efficient and relies only on addition and scalar multiplication to extract a compact, distinct, and robust signature based on the global features on the content. This simplicity of computation enables the signature computation and comparison equivalently on the encrypted content. Though simple, it is still robust to common image manipulations that are most likely to happen during the sharing of videos over social media. Another advantage of the proposed scheme is that the hash extracted from encrypted content is in plaintext domain. It can be directly compared for a match with the database content signatures, whether they may be kept encrypted or unencrypted. For the entire protocol, the server requires only one round of pre-interaction with the client at the very start where some encrypted metadata needs to be sent by the client apart from the encrypted content. For a false positive rate of $1$ in $10$ million, the proposed scheme's true positive rate remains high at $93.5\%$. We envision deploying this technology can tackle the problem of revenge porn as it resolves the people's primary concern in the solution proposed by Facebook, where they have to compromise with their privacy to avail of the possible solution.

\section{Paillier Cryptosystem}
\label{sec:paillier}

 The Paillier cryptosystem is a partially homomorphic and asymmetric encryption scheme~\cite{paillier1999public}. The encryption/decryption is done using a public-private key pair. \\

A plaintext message $m$ can be encrypted to ciphertext $c$ as follows:
\begin{eqnarray}
c ~=~ E(m,r;g,n) ~=~ \mbox{mod}(g^m \times r^{n}, n^{2}),
\label{eqn:encrypt}
\end{eqnarray}
where $r$ is a random integer satisfying $0<r<n$. Incorporating this random value ensures that the same plaintext is encoded as different ciphertexts under the same public key.

The public key ($n$, $g$) is generated using two large prime numbers $p$ and $q$ as follows: 
\begin{eqnarray}
n & = & p \times q \\
g & = & n + 1
\end{eqnarray}

A ciphertext message $c$ is decrypted as follows:
\begin{eqnarray}
D(c;\lambda,\mu,n) & = & \mbox{mod} \left( \frac{\mbox{mod}(c^\lambda,n^2) - 1}{n} \times \mu, n \right).
\end{eqnarray}

The private key ($\lambda$, $\mu$) is computed as:
\begin{eqnarray}
\lambda & = & \mbox{lcm}(p-1, q-1) \\
\mu     & = & \mbox{mod}(\alpha,n),
\end{eqnarray}
where $\mbox{lcm}(\cdot)$ denotes the least common multiple operator, $\mbox{mod}(\cdot)$ is the modulo operator, and the extended Euclidean algorithm computes $ \alpha$. Specifically, $\gamma$ is first computed as follows:
\begin{eqnarray}
\gamma & = & \mbox{gcd} \left( \frac{\mbox{mod}(g^\lambda,n^2)-1}{n}, n \right),
\label{eqn:gcd}
\end{eqnarray}
where $\mbox{gcd}(\cdot)$ is the greatest common divisor operator. The B\'ezout coefficients $\alpha$ and $\beta$ satisfy:

\begin{eqnarray}
\gamma & = & \alpha \times \frac{\mbox{mod}(g^\lambda,n^2)-1}{n} + \beta \times n
\label{eqn:modinv}
\end{eqnarray}

The homomorphic properties satisfied by the Paillier cryptosystem are addition and scalar multiplication.
\begin{itemize}
	\item The sum of two plaintexts messages $m_1$ and $m_2$ is equal to the decrypted product of two ciphertexts $c_1$ and $c_2$ under the appropriate modular arithmetic:
	\begin{eqnarray}	
	 \mbox{mod}(m_1+m_2,n) = D(\mbox{mod}(c_1 \times c_2, n^2);\lambda, \mu, n)
	\label{eqn:homomorphic1}
	\end{eqnarray}
	\item The product of a scalar $s$ and a plaintext $m_1$ is equal to the decrypted exponentiation of ciphertext by a scalar value under the appropriate modular arithmetic:
	\begin{eqnarray}	
	 \mbox{mod}(s \times m_1,n)= D(\mbox{mod}(c_1^s, n^2);\lambda, \mu, n)
	\label{eqn:homomorphic2}
	\end{eqnarray}
\end{itemize}

We will describe a robust video hashing scheme (extraction and comparison) that requires only addition and scalar multiplication in the plaintext domain and can be fully transformed in the encrypted domain.

\section{Robust Video Hashing}
\label{sec:videohash}
A brute-force approach to video hashing is to simply hash every individual frame and concatenate the frame hashes into one long hash for comparison against a reference hash. Although attractive for its simplicity, this approach is not particularly desirable. In particular, the resulting hash will be prohibitively large: At a standard $30$ frames/second, and even modest three-minute video will yield $5,400$ individual hashes, each of some fixed length based on the hashing algorithm. Such a large hash will lead to significant computational and bandwidth demands. Given the highly redundant nature of most videos, this simple but inefficient approach is unnecessary. In the proposed method, we adopt selectively hashing a particular set of video frames contrary to hashing every video frame. Only those video frames that provide new identifiable information not previously seen in the video stream are incorporated to contribute to the video hash. The basic framework of the video hashing is depicted in Fig.1. It has four main phases: the selection of content frames, pre-processing of frames,  computation of video hash, and comparison of video hashes. Each phase of the video hashing pipeline consists of tightly coupled sub-phases that altogether achieve a specific objective and facilitates the execution of the next phase. The  details are as follows: 

\begin{figure*}
	\centering
	\includegraphics[scale=0.45]{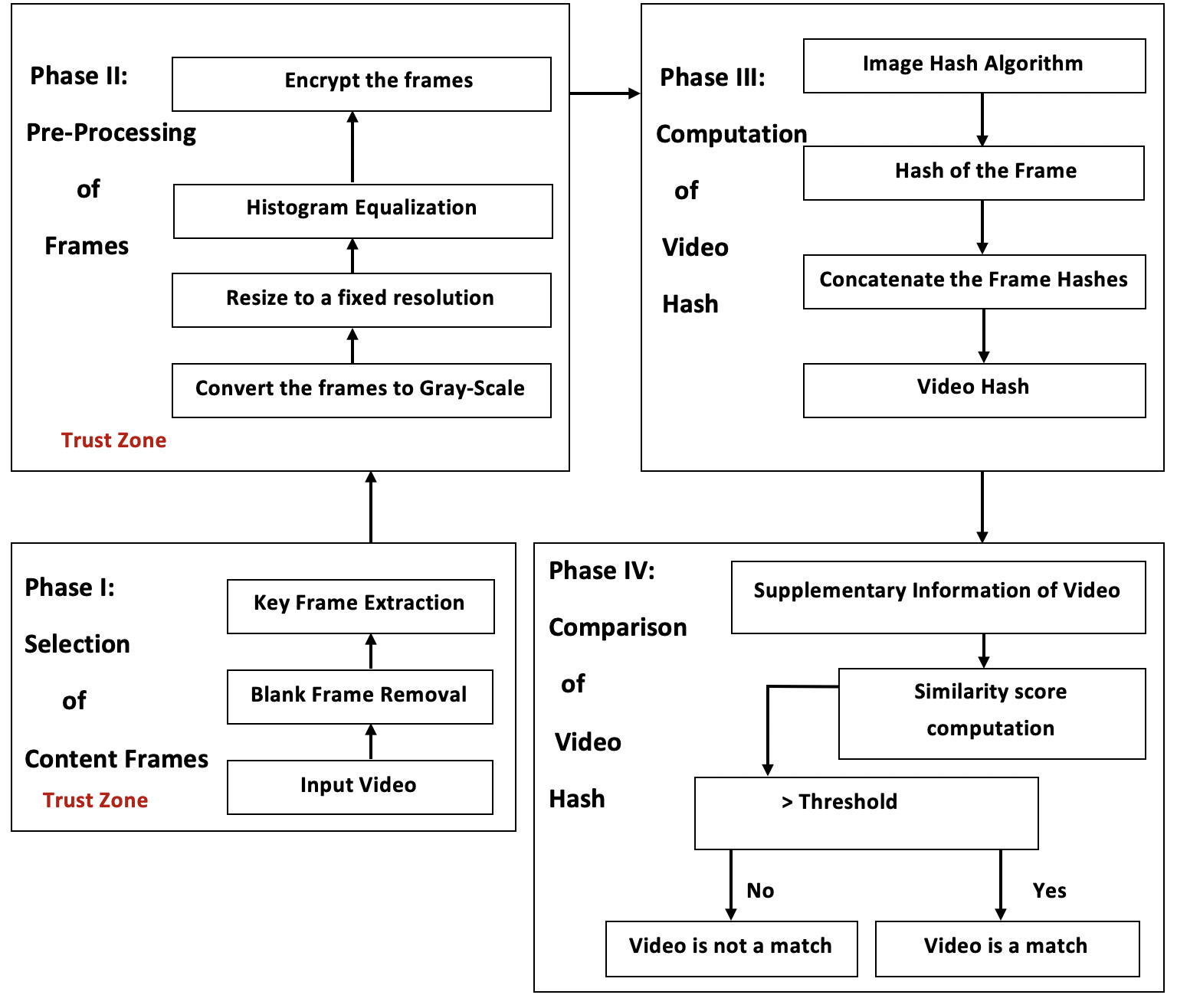}
	\caption{Overview of Video Hashing}
	\label{fig:roc-video}
\end{figure*}

\subsection{Selection of Content Frames}
A video hash must be a compact and distinct signature that can uniquely identify a video amongst billions and trillions of videos floating on the internet. This phase eliminates the redundant information in a video by dropping the blank frames or similar information content frames because that leads to an unnecessary increase in length without fetching the advantage of added information. As video length increases, the cost of processing and storing a video hash increases at a tremendous rate.
This phase has blank frame removal, and keyframe extraction as its two sub-phases. 

\subsubsection{Blank Frame Removal}
In a video, the frames that do not have much information or content variation can be considered blank frames. Mostly, these frames represent homogeneous surfaces like walls, blackboards, or natural scenic stretch like frames contributing to sky appearance, water bodies, etc. Removal of such frames will incur no information loss of a video and facilitate the next keyframe extraction phase.  

\subsubsection{Key Frame Extraction} 
Keyframes need to be extracted from a bucket of content frames after the removal of the blank frames. Even at a standard rate of $30$ frames/second, much redundant information is captured to give a sense of smooth motion to the human eyes. This increment of information is directly proportional to the increase in the frame rate. 	In key frame extraction, a consecutive frame is considered as a keyframe only if there is a significant addition to the information recorded for the previous keyframe. Any standard distance metric can be employed for the frame hashes to extract the keyframes. If the distance exceeds a threshold value, then the successive frame is kept as keyframe else discarded. At this point, some supplementary information like the number of frames dropped in between two keyframes, indices of keyframes, percentage of keyframes as a function of the total number of content frames, etc. can be recorded. This process is done for the entire video. 

\subsection{Pre-processing of Frames}
To further enhance the computational complexity of a video hash, some pre-processing to the extracted keyframes can be performed. The basic pre-processing includes converting the frames to gray-scale, fixing a resolution for the frames, and histogram equalization of the fixed-size frames.   
Instead of processing individual color channels, it is faster to process the gray-scale version of a frame. Resizing to a fixed resolution (practically feasible to process) enhances the processing speed. Histogram equalization can take care of the varying illumination conditions in which the video was shot. So, the video similarity score doesn't get affected by this factor and is mainly focused on capturing the information about the content. 


\subsection{Computation of Video Hash}
Once the keyframes are extracted and pre-processed, a standard image hashing algorithm can be employed to extract the hash of these frames. The hashes of the frames are concatenated in a temporal order to obtain the final video hash. The length of a video hash is a variable quantity and depends on many factors like the length of the video snippet, the number of keyframes, the presence of blank frames in the original video, etc.  

\subsection{Comparison of video hashes}
 
Unlike image hashes that are guaranteed to be the same length, video hashes can be of arbitrary length. As a result, two video hashes of different lengths can not be compared using the traditional distance measures like L1-norm, Euclidean distance, Minkowski distance, etc. as used to compare image hashes. Variable-length hashes are a bit of a nuisance, but also create an opportunity. 

We utilize the longest common substring (LCS) (not to be confused with the longest common sub-sequence algorithm) \cite{DanGusfield} to measure the similarity of two variable-length video hashes. By way of intuition, the LCS of the two strings ``ABABACABBC" and ``ABACABACBBCA" is six because the longest common string shared by these strings is ``ABACAB." Note that these strings also have the substring ``BBC" in common, but this is shorter than the substring of length six. Given two strings of length $m$ and $n$, the LCS can be found efficiently using dynamic programming with a run-time complexity of $O(mn)$. The advantage of using LCS to compare two hashes is that it allows us to find not just matching videos but also video segments that are extracted or video segments that are embedded within a larger video (e.g.,~a video compilation). Within this LCS computation, two hashes, corresponding to a single keyframe, are of course, of the same length as the image hash algorithm used for its computation. However,  LCS alone was not able to capture the similarity of two matching videos.

To take care of the situation, we incorporated two factors to determine when to increase the LCS count for two matching videos. First is, of course, the hash values for the corresponding keyframes of the two videos must match. This could be measured using any standard distance metric. We adopted the L1-norm to compute the difference of the hash values for the corresponding frames for simplicity. It must fall below a threshold to indicate a considerable similarity of the frames.
The second most important factor that we incorporated was considering the number of video frames dropped in between the consecutive keyframes. For two videos to be similar, the temporal information needs to be matched on this basis. If the difference between the number of dropped frames of the corresponding keyframes falls below a threshold value, then only it signifies the temporal similarity. So, whenever both the aforementioned conditions are satisfied, the LCS count is incremented. The increment is done by the average of the dropped frames for the corresponding keyframes. This is repeated for matching the entire video sequence. The final similarity score for the videos is this LCS value.


\section{Transformation in the Encrypted Domain}
\label{sec:encrypted}
The threat to privacy is a significant concern for almost all applications. It has fueled the research in the encrypted domain. We look for video hash that can be transformed in the encrypted domain. 
It is possible only when the computations are based on properties supported by homomorphic encryption schemes. In the proposed video hash framework, this is very much achievable as the image hash algorithm used is the robust homomorphic image hashing algorithm proposed in \cite{Singh_2019_CVPR_Workshops}. The computations involved in the hash extraction are limited to addition and scalar multiplication. These computations are supported by Paillier homomorphic encryption scheme.

To enable the video hash in the encrypted domain, the two phases: the selection of content frames and pre-processing of frames must be done is a trusted zone like the ARM TrustZone. The TrustZone provides a secure environment by providing a secure operating system, secure storage, and a secure set of applications. The processor core in the TrustZone is divided into two virtual cores referred to as worlds: normal world and secure world. Access to the secure world is highly regulated. Hence, sensitive applications can operate in this secure environment. Non-secure applications can be executed in the normal world. Both the secure world and the normal world are separated by hardware extensions. 

At the end of the second phase, the pre-processed frames can be encrypted using the homomorphic Paillier encryption. Once the frames are encrypted, the last two phases of the video hash algorithm can work outside the TrustZone. In phase 3, the video hash is obtained via the concatenation of frame hashes in temporal order. As the hash of the encrypted frames is extracted using the image hash algorithm proposed in \cite{Singh_2019_CVPR_Workshops}, the hash is in plaintext domain. The video hash is the concatenation of these plaintext hashes, and hence, it will also be in plaintext domain. 

Phase 4 of the video hash framework where comparison of video hashes is made, will work on top of these plaintext video hashes. The supplementary information of a video needed for computation of similarity score comprises of the count of the dropped frames in between the keyframes. This information can be provided in the plaintext domain itself as no substantial leakage of information about the original video sequence can take place from this particular information. Hence, video hash computation is feasible in the encrypted domain. This provides much potential for addressing issues of privacy-preserving applications and combating illegal activities happening on the encrypted channels.

\section{Experimental Validation and Robustness}
\label{sec:exp}
We constructed a dataset of approximately $12000$ videos ranging in length from $2$ to $6$ seconds. Using the plaintext video hashing described above, we hashed all of these videos and computed a pairwise comparison of all videos to remove any duplicates. Any video pairs with an LCS below a specified threshold were manually reviewed to determine if they were duplicates. One of the pairs of any confirmed duplicates was removed from our dataset, yielding a final video count of $11,278$.

In order to determine the robustness of our video hash to common image manipulations, we generated $5500$ variations of each video with varying and random amounts of gamma correction $I^\gamma(x,y)$ with $\gamma \in [0.5,2]$, additive Gaussian noise with SNR in the range $[15,60]$ (dB), MPEG compression with quality in the range $[2,23]$. The quality ranges from  0 to 51, where 0 is lossless, 23 is the default, and 51 is the worst quality possible. Scaling is done in the range $[0.25,1]$. 

These manipulations were applied to the video prior to the pre-processing stage. We computed the LCS distance between the original video and each of its variants, yielding a total of over $62$ million (``similar") comparisons. We also computed the LCS distance between the hash of each of the original videos and the remaining $11278-1$ videos, yielding a total of over $63$ million (``different") comparisons. 

\begin{figure}[t]
	\centering
	\includegraphics[scale=0.23]{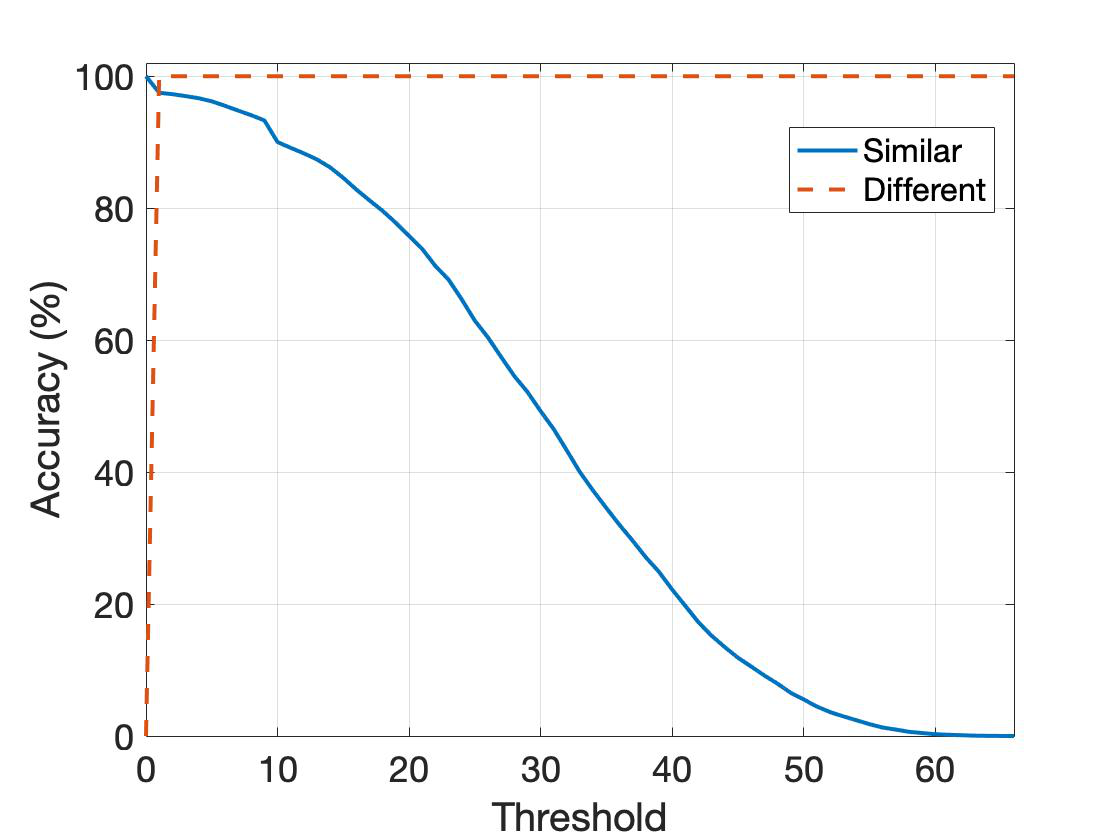}
	\caption{Accuracy of classifying similar (solid-blue) and different (dashed-red) videos as a function of the threshold on the longest common substring (LCS) between two hashes. The cross-over point is $97.5\%$. See also Figure ~\ref{fig:gamma_snr_mpeg-video}}. 
	\label{fig:roc-video}
\end{figure}
\begin{figure}[ht]
	\begin{center}
		\includegraphics[scale=0.19]{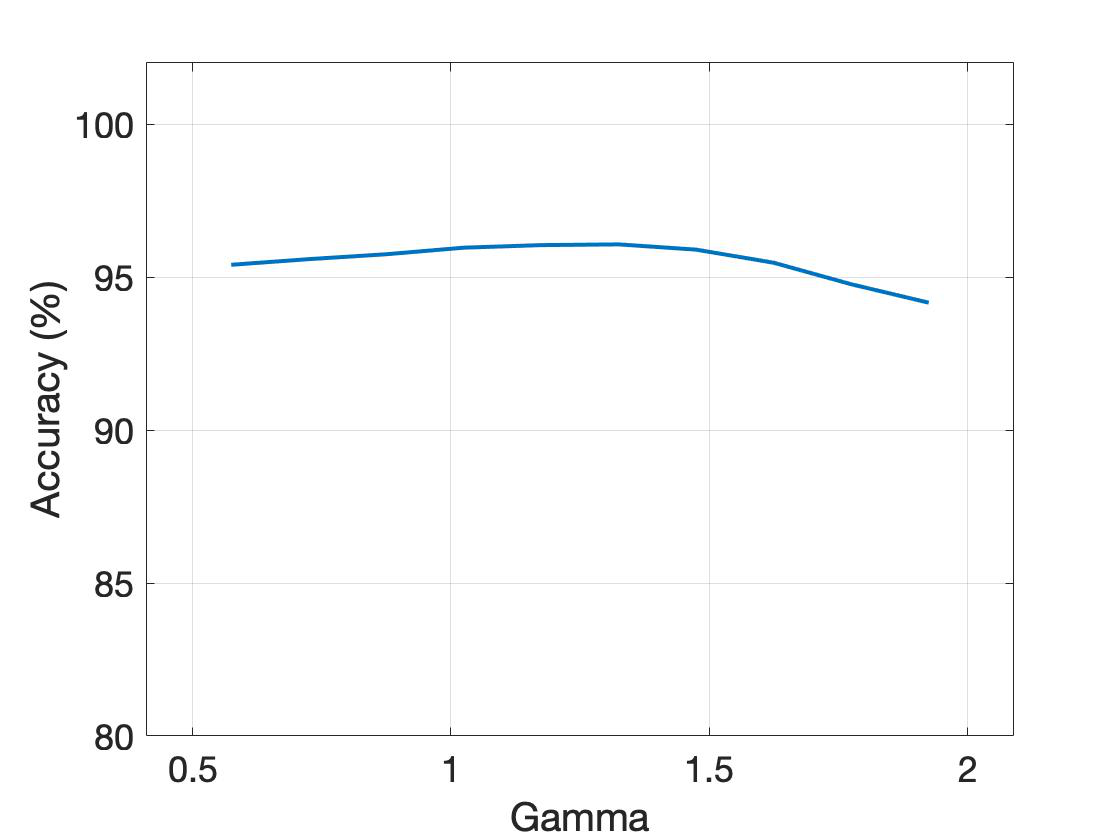}
		\includegraphics[scale=0.19]{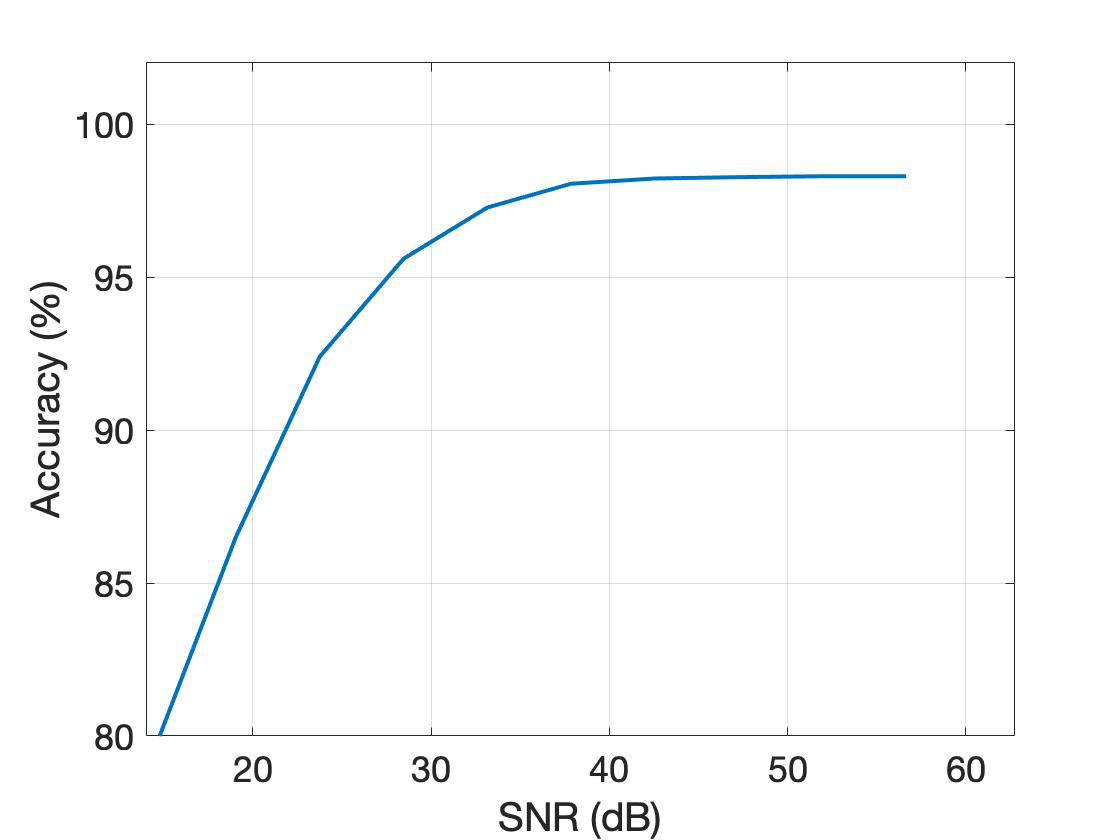}
		\includegraphics[scale=0.19]{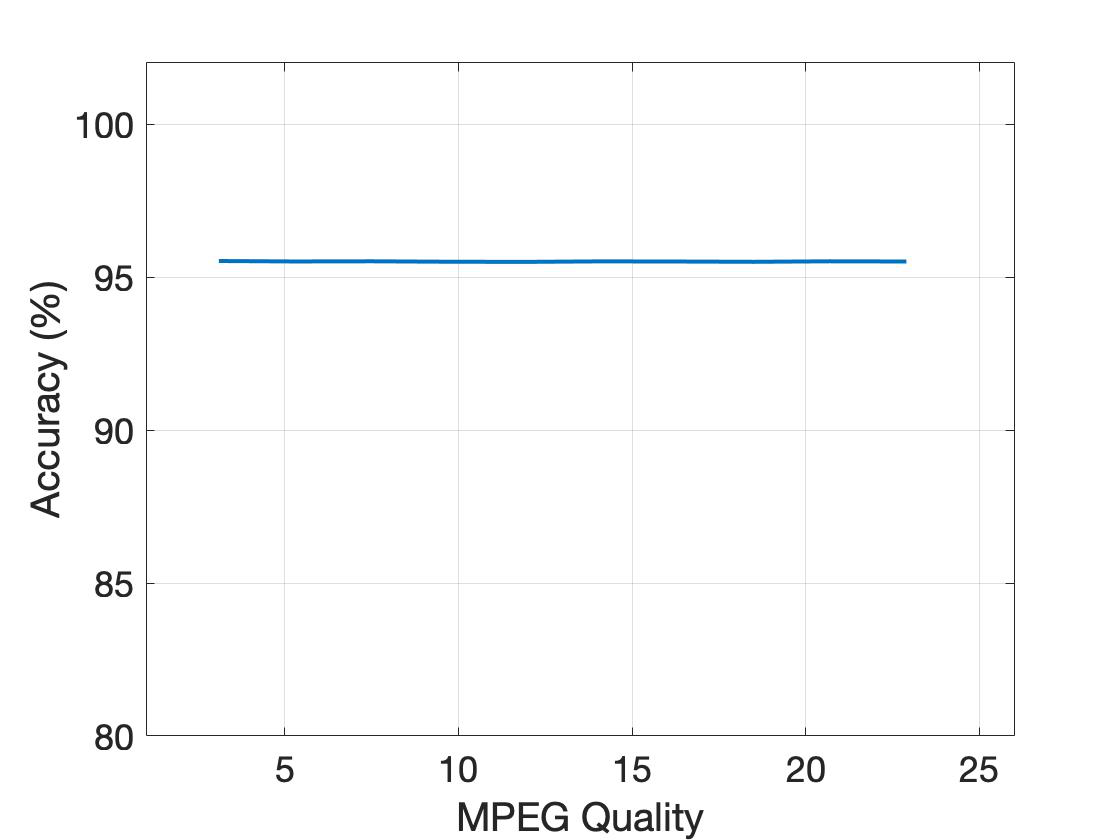}
		\includegraphics[scale=0.19]{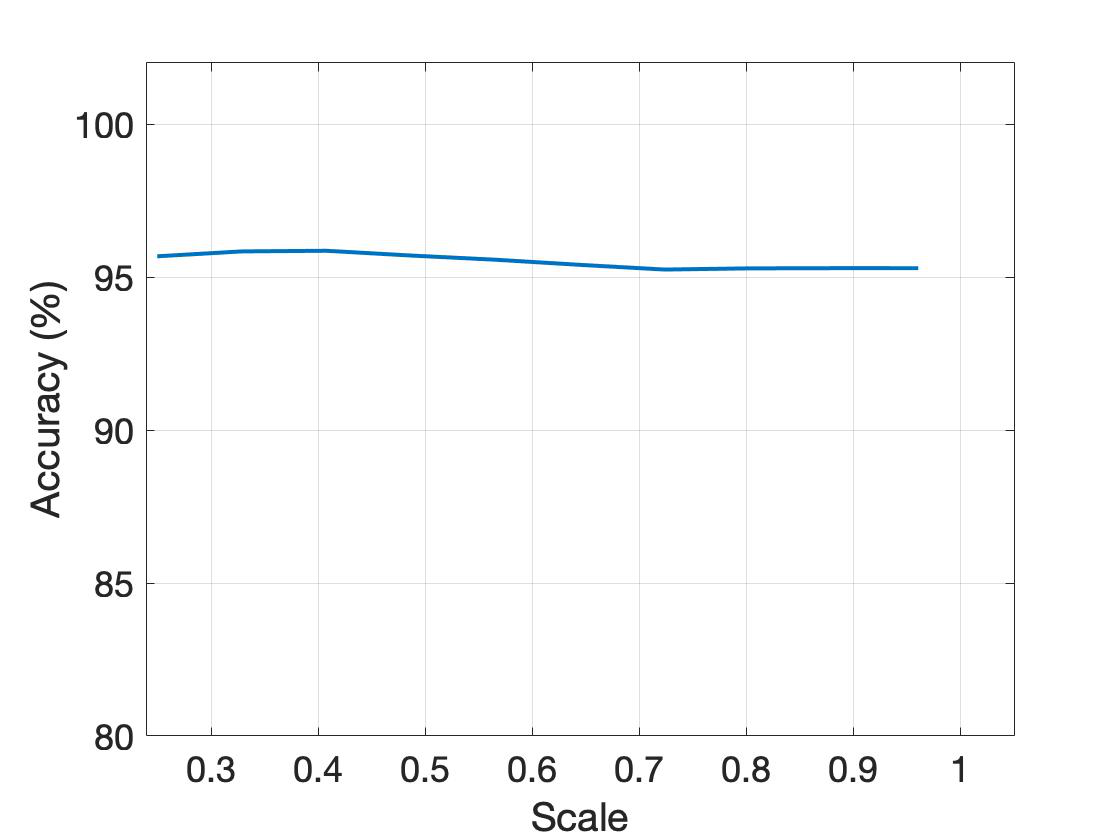}
	\end{center}
	\caption{Accuracy of classifying similar videos as a function of gamma correction, additive noise, MPEG quality, and scaling, for a fixed false alarm rate of $0.001\%$.}
	\label{fig:gamma_snr_mpeg-video}
\end{figure}

A robust video hash should have relatively large ``similar" distances and small ``different" distances. The receiver operating characteristic (ROC) curve in Fig.~\ref{fig:roc-video} reports the accuracy of correctly classifying ``similar" (solid-blue) and ``different" (dashed-red) videos as a function of the LCS threshold. The cross-over point of these curves is at $97.5\%$ accuracy. With a false positive rate (incorrectly identifying different videos as the same) of $0.01\%$, and $0.001\%$, the true positive rate (correctly identifying similar videos as similar) is $97.5\%$,  and $95.5\%$. At a false positive rate of $1$ in $10$ million, the true positive rate remains relatively high at $88.4\%$.

Shown in Fig.~\ref{fig:gamma_snr_mpeg-video}, from top to bottom, is the true positive rate (with a fixed false-positive rate of $0.001\%$) as a function of the amount of gamma correction, additive noise,  MPEG quality, and scaling. Each data point in each panel corresponded to all videos with the specified distortion and integrated across all other distortions. From top to bottom, we see that: (1) the sensitivity to gamma correction is asymmetric with slightly lower robustness to gamma values greater than $1.0$; (2) the hash is robust to SNR larger than $25$ dB; and (3) the hash is robust to a range of compression qualities and scalings. 


\subsection{Efficiency}
The experiments were conducted on a machine running macOS with an Intel Core i5, 2.3 GHz. MatLab was used for the implementation of plaintext video hash, whereas the ciphertext version was done in Java. The video length and pre-processing steps do affect the run-time efficiency of the algorithm. For our dataset of 12K videos of duration ranging from 2 seconds to 6 seconds,  the average run-time for extraction of plaintext video hash was 0.05 seconds and 15.3 seconds for the ciphertext version.

\section{Discussion}
\label{sec:discussion}
Privacy-preserving solutions that can help secure the privacy of outsourced data floating on social media sites but still help combat propagation of illegal content is the need of the hour. This balance can be achieved by exploiting the encryption schemes' homomorphic properties and focusing on the design of efficient algorithms based on the simple computations. We proposed a video hash algorithm in this paper and discussed for possible transformation in the ciphertext domain.
In the future, an integrated solution incorporating the other modalities like text and audio can be coupled together with images and videos to give a complete solution.

\bibliographystyle{IEEEtran}
\bibliography{ref}

\end{document}